\documentstyle[aps,floats,prd,psfig]{revtex}

\newcommand{\ApJL}{Astrophys. J. Lett.}
\newcommand{\ApJ}{Astrophys. J.}
\newcommand{\PRL}{Phys. Rev. Lett.}
\newcommand{\PRD}{Phys. Rev. D}
\newcommand{\MNRAS}{Mon. Not. Royal Ast. Soc.}
\newcommand{\AsAs}{Astron.  Astrophys.}
\newcommand{\aut}[2]{{#2.\ #1}}
\newcommand{\refs}[6]{#2, {\bf #3} {#4} (#5)}

\newcommand{\amp}{and }

\newcommand{\tot}{{\rm t}}

\newcommand{\cmb}{\Theta}
\newcommand{\s}{{\rm s}}
\newcommand{\n}{{\rm n}}
\newcommand{\vecl}{{\bf l}}
\newcommand{\vecla}{{{\bf l}_1}}
\newcommand{\veclb}{{{\bf l}_2}}
\newcommand{\veclc}{{{\bf l}_3}}
\newcommand{\vecld}{{{\bf l}_4}}
\newcommand{\vecle}{{{\bf l}_5}}
\newcommand{\veclf}{{{\bf l}_6}}
\newcommand{\intl}[1]{\int {d^2 {\bf l}_#1 \over (2\pi)^2}}

\newcommand{\bfl}{{\mathbf{l}}}
\newcommand{\dirac}{{\rm D}}

\newcommand{\pp}{{\phi\phi}}

\newlength{\tskip}
\newlength{\colwidth}

\newcommand{\beq}{\begin{equation}}
\newcommand{\eeq}{\end{equation}}
\newcommand{\beqa}{\begin{eqnarray}}
\newcommand{\eeqa}{\end{eqnarray}}
\newcommand{\bn}{\hat{\bf n}}

\newcommand{\len}{\phi}

\begin{document}
%@twocolumnfalse\endcsname

\title{Weak Lensing of the CMB: Cumulants of the Probability Distribution Function}
\author{Michael Kesden, Asantha Cooray, and Marc Kamionkowski}
\address{
Theoretical Astrophysics, California Institute of Technology,
Pasadena, California 91125\\}

%\date{To be submitted to Phys. Rev. D. --- October 2001}

\maketitle

%------------------------------------------------------------------------------

\begin{abstract}
We discuss the real-space moments of temperature anisotropies in the cosmic
microwave background (CMB) due to weak gravitational lensing by intervening
large-scale structure.  We show that if the probability distribution function
of primordial temperature anisotropies is
Gaussian, then it remains unchanged after gravitational lensing.
With finite resolution, however, non-zero higher-order cumulants are generated
both by lensing autocorrelations and by cross-correlations between the lensing
potential and secondary anisotropies in the CMB such as the Sunayev-Zel'dovich
(SZ) effect.  Skewness is produced by these lensing-SZ correlations, while
kurtosis receives contributions from both lensing alone and
lensing-SZ correlations.  We show that if the projected lensing potential is
Gaussian, all cumulants of higher-order than the kurtosis vanish.
While recent results raise the possibility of
detection of the skewness in upcoming data, the kurtosis will likely remain
undetected.
\end{abstract}

%]

%------------------------------------------------------------------------------
% User-supplied List of keywords.

%\pacs{PACS numbers: 98.80.Es,95.85.Nv,98.35.Ce,98.70.Vc
%\hfill}
%]

%------------------------------------------------------------------------------

\section{Introduction}
Weak gravitational lensing deflects the paths of cosmic microwave background
(CMB) photons propagating from the surface of last scattering.
One result of this lensing is
the transfer of power from large angular scales associated with acoustic-peak
structures to small angular scales
in the damping tail of the anisotropy power spectrum \cite{Blaetal87,Hu00}.
This transfer only results in a few-percent modification of the power
associated with the acoustic-peak structure, and
the increase in power along the damping tail is significantly smaller than that
generated by
secondary anisotropies due to reionization \cite{Coo02}.  
To indentify the effect of gravitational lensing on CMB data, 
it is necessary to consider signatures beyond that in the angular power
spectrum of temperature fluctuations.
The existence of non-vanishing higher order cumulants is one such non-Gaussian
signature lensing can generate.
%(e.g., \cite{SpeGol99,CooHu00,Hu01,Benetal00}).

Since gravitational lensing conserves surface brightness, CMB fluctuations from
lensing are at the second order in temperature fluctuations and result in
non-Gaussian behavior through non-linear mode coupling. Though
lensing alone does not lead to a three-point
correlation function, the correlation between lensing and other secondary
anisotropies can lead to such a contribution. This 
three-point correlation has been widely discussed in the literature in terms of
its Fourier-space analogue, the bispectrum 
\cite{SpeGol99}.  Weak lensing of the
primary anisotropies can produce a four-point correlation
due to its non-linear mode-coupling nature \cite{Bern97,Zal00,Coo02a}, as can
correlations between
lensing and secondary effects \cite{Coo02a}.
When probed appropriately through quadratic statistics such as the power spectrum
of the squared-temperature map,
the trispectrum due to lensing alone can be used for
a model-independent recovery of the projected mass distribution out to the last
scattering surface \cite{Hu01b,CooKes02}.
Though these statistics have been shown to be interesting and potentially
detectable, measurement of these Fourier-based statistics is challenging
and techniques are still underdeveloped for this purpose.

Here, we discuss real-space moments of the lensed CMB temperature anisotropies.
Real-space statistics
are easily measurable from data. The only drawbacks
are that
they are unlikely to be optimal and only provide limited knowledge of the full
non-Gaussian aspect of the temperature distribution.
The first attempts to measure non-Gaussianity in the COBE data relied on
real-space cumulants \cite{Hinetal95}, as will attempts using data from its
successor experiments such as MAP and Planck.
This motivates our emphasis here on the real-space cumulants such as the
skewness and kurtosis; we make several remarks on higher-order cumulants as
well.  

As part of this calculation, 
we extend a previous discussion of the kurtosis due to lensing in
Ref. \cite{Bern97}
and also consider effects related to correlations between lensing and
secondary effects such as the Sunyave-Zel'dovich (SZ; \cite{SunZel80}) effect.

Real-space moments can be derived from the one-point probability distribution
function (PDF) of temperature fluctuations, and can conversely be used to
constrain the form of this function.  In the case of infinite angular
resolution, we conclude that lensing does not modify the PDF of 
temperature anisotropies produced at the last scattering surface, which is a
reflection on the fact that lensing does not create new power but rather
transfers power from large to small angular scales.
The higher-order moments are only generated in a temperature map by
finite-resolution effects such as beam smoothing
introduced either experimentally or artificially by explicit filtering.

The paper is organized as follows.
In \S~\ref{sec:lensing}, we introduce formalism concerning the weak-lensing
approximation and define the bispectrum, trispectrum, and corresponding
higher-order quantities.  The bispectrum and trispetrum induced in the CMB by
lensing and secondary anisotropies are derived in \S~\ref{sec:PBT}, and some
remarks are made concerning higher-order cumulants as well.  The nonzero
bispectrum and trispectrum yield a skewness and kurtosis respectively in the
one-point distribution function of the CMB as shown in \S~\ref{sec:S&K}.
We refer the reader to Ref. \cite{CooKes02} for additional details related to
the effect of lensing on CMB anisotropies. Though we present a general
discussion, we illustrate our results in \S~\ref{sec:results} using
the currently favored $\Lambda$CDM cosmological model with $\Omega_b=0.05$,
$\Omega_m=0.35$, $\Omega_\Lambda=0.65$, $h$=0.65, and $\sigma_8 = 0.9$.
Results for a model with $\sigma_8 = 1.2$ as suggested by CBI are also
considered.

\section{Lensing Contribution to CMB Fluctuations}
\label{sec:lensing}

In order to derive the effects of weak lensing on the CMB, we
follow  Refs. \cite{Hu00,Zal00} and adopt a flat sky approximation. 
As discussed in prior papers \cite{Hu00,SpeGol99},  weak lensing 
remaps temperature through angular deflections along the photon path:
\begin{eqnarray}
\tilde \cmb(\bn) & = &  \cmb(\bn + \nabla\len) \nonumber\\
        & = &
\cmb(\bn) + \nabla_i \len(\bn) \nabla^i \cmb(\bn) + {1 \over 2} \nabla_i \len(\bn) \nabla_j \len(\bn)
\nabla^{i}\nabla^{j} \cmb(\bn)
+ \ldots  \, .
\end{eqnarray}
Here, $\cmb(\bn)$ is the unlensed primary component of the CMB in direction
$\bn$ at the last scattering surface, $\tilde \cmb(\bn)$ is the lensed map,
$\len(\bn)$ is the projected gravitational potential, and $\nabla\len$ is the
lensing deflection angle.
It should be understood that in the presence of low-redshift
contributions to CMB fluctuations resulting from large-scale 
structure, the total map includes secondary contributions which we denote by
$\cmb^\s(\bn)$.
Since the weak-lensing deflection angles $\nabla\len$ also trace the
large-scale structure at low redshifts, 
secondary effects which are first order in density
fluctuations correlate with the lensing deflection angles.
These secondary effects include the integrated Sachs-Wolfe
(ISW; \cite{SacWol67}) and the SZ \cite{SunZel80} effects \cite{SpeGol99}.
In all real cases, a noise component denoted by $\cmb^\n(\bn)$ due to finite
experimental sensitivity must be included as well.
Thus the total observed CMB anisotropy will be
$\cmb^\tot(\bn) = \tilde \cmb(\bn) + \cmb^\s(\bn) + \cmb^\n(\bn)$.  In the
following discussion, secondary anisotropies $\cmb^\s(\bn)$ will be neglected
until subsection \ref{SS:sec} while the effects of instrumental noise
$\cmb^\n(\bn)$ on the PDF are discussed in \S~\ref{sec:S&K}.

Taking  the Fourier transform, as appropriate for a flat sky, we write
\begin{eqnarray}
\tilde \cmb(\vecl)
&=& \int d \bn\, \tilde \cmb(\bn) e^{-i \vecl \cdot \bn} \nonumber\\
&=& \cmb(\vecl) - \intl{1'} \cmb(\vecl') L(\vecl,\vecl')\,,
\label{eqn:thetal}
\end{eqnarray}
where
\begin{eqnarray}
\label{eqn:lfactor}
L(\vecl,\vecl') &=& \len(\vecl-\vecl') \,
\left[ (\vecl - \vecl') \cdot \vecl' \right]
+{1 \over 2} \intl{1''} \len(\vecl'') \\ &&\quad
\times \len^*(\vecl'' + \vecl' - \vecl) \, (\vecl'' \cdot \vecl')
%\nonumber\\ &\times&
                \left[ (\vecl'' + \vecl' - \vecl)\cdot
                             \vecl' \right] \,.  \nonumber
\end{eqnarray}

We define the power spectrum, bispectrum, trispectrum and the $n$-point
correlator in Fourier space in the usual way:
\begin{eqnarray}
\left< \tilde \cmb(\bfl_1) \tilde \cmb(\bfl_2)\right> &=&
        (2\pi)^2 \delta_\dirac(\bfl_{12}) \tilde C_{l_1}^\cmb\,,\nonumber\\
\left< \tilde \cmb(\bfl_1) \ldots
       \tilde \cmb(\bfl_3)\right>_c &=& (2\pi)^2 \delta_\dirac(\bfl_{123})
        \tilde B^\cmb(\bfl_1,\bfl_2,\bfl_3)\,. \nonumber \\
\left< \tilde \cmb(\bfl_1) \ldots
       \tilde \cmb(\bfl_4)\right>_c &=& (2\pi)^2 \delta_\dirac(\bfl_{1234})
        \tilde T^\cmb(\bfl_1,\bfl_2,\bfl_3,\bfl_4)\,. \nonumber \\
\left< \tilde \cmb(\bfl_1) \ldots
       \tilde \cmb(\bfl_n)\right>_c &=& (2\pi)^2 \delta_\dirac(\bfl_{1 \ldots n})
        \tilde T_{n}^\cmb(\bfl_1, \ldots ,\bfl_n)\,. \nonumber \\
\end{eqnarray}
where $\bfl_{1 \ldots n} \equiv \bfl_1 + \ldots + \bfl_n$, and the subscript
$c$ denotes the connected portion of the correlation function.
We make the assumption that primary anisotropies at the last scattering surface are Gaussian implying that all cumulants higher than the power spectrum vanish: 
$\left< \cmb(\bfl_1) \ldots \cmb(\bfl_n)\right>_c = 0$, when $n > 2$.

The $n$th cumulant of the temperature anisotropies is defined in the usual
manner,
\begin{equation}
T^n(\theta) = \int \frac{d^2\vecl_1}{(2\pi)^2} \ldots \frac{d^2\vecl_n}{(2\pi)^2} 
\left< \cmb^\tot(\bfl_1) \ldots \cmb^\tot(\bfl_n)\right>_c W(l_1 \theta) \ldots W(l_n \theta),
\label{eqn:mom}
\end{equation}
where $\theta$ is the smoothing scale of the map from which the cumulants are
determined, and $W(l \theta)$ is the
smoothing window function. We will use Gaussian window functions throughout this paper. In general,
the finite resolution of real CMB anisotropy experiments induces Gaussian
smoothing at the angular scale of the experimental beam size. For infinite
resolution, we take $\theta \rightarrow 0$ such that
$W(l \theta) \rightarrow 1$.

\section{Power spectrum, Bispectrum and Trispectrum}
\label{sec:PBT}

Using the formalism introduced in the previous Section, we can calculate the
moments of the CMB fluctuations generated by lensing assuming Gaussian
fluctuations at the surface of last scatter.
The power spectrum for the lensed map is \cite{Blaetal87,Hu00}
\begin{eqnarray}
\tilde C_l^\cmb &=& \left[ 1 - \intl{1}
C^{\phi\phi}_{l_1} \left(\vecl_1 \cdot \vecl\right)^2 \right]	\, 
 	                        C_l^\cmb
        + \intl{1} C_{| \vecl - \vecl_1|}^\cmb C^{\phi\phi}_{l_1}
                [(\vecl - \vecl_1)\cdot \vecl_1]^2  \, .
\label{eqn:ttflat}
\end{eqnarray}

The variance, or the second moment of the temperature, can be obtained following Eq.~(\ref{eqn:mom})
\begin{equation}
\sigma^2(\theta) = \int \frac{d^2\vecl}{(2\pi)^2} \tilde C_l^\cmb W^2(l \theta) \, .
\end{equation}
Substituting Eq.~(\ref{eqn:ttflat}) in here, we find that in the case of
infinite resolution ($W(l\theta)=1$), the variance of the lensed temperature
map coincides with that of the unlensed map. Thus, as expected,
lensing conserves the total power associated with the temperature fluctuations. This is consistent with
our basic expectation that lensing only results in a transfer of power from large angular scales to
small angular scales. With finite resolution at levels considered here, the
variance of the lensed temperature field
differs from that of the unlensed field by a few percent at most.

We will now discuss higher-order correlations of temperature due to
gravitational lensing. We consider first contributions due to lensing alone,
and then discuss additional contributions created by lensing-secondary
correlations.

\subsection{Lensing Correlations}

We will first discuss the temperature bispectrum and show that it is zero in
the absence of secondary anisotropies. 
To understand why there is no contribution to the bispectrum, consider the 
moments involving  three temperature terms in Fourier space:
\begin{eqnarray}
&&\left< \tilde \cmb(\bfl_1) \tilde \cmb(\bfl_2)
       \tilde \cmb(\bfl_3)\right>_c =  \nonumber \\
&&\Big< \left( \cmb(\vecla) - \intl{1'} \cmb(\vecla') L(\vecla,\vecla') \right)
\left( \cmb(\veclb) - \intl{2'} \cmb(\veclb') L(\veclb,\veclb') \right)
\left( \cmb(\veclc) - \intl{3'} \cmb(\veclc') L(\veclc,\veclc') \right)\Big>
\nonumber \\
&& \quad =  \Big< \cmb(\vecla) \cmb(\veclb)  \cmb(\veclc) \Big>
- \Big< \cmb(\vecla) \cmb(\veclb) 
	\left( \intl{3'} \cmb(\veclc') L(\veclc,\veclc')  \right) \Big> + {\rm Perm.}  \nonumber \\
&+& \Big< \cmb(\vecla) \left( \intl{2'} \cmb(\veclb') L(\veclb,\veclb')  \right)
	\left( \intl{3'} \cmb(\veclc') L(\veclc,\veclc')  \right) \Big> + {\rm Perm.} \nonumber \\
&-& \Big< \left( \intl{1'} \cmb(\vecla') L(\vecla,\vecla')  \right)
	\left( \intl{2'} \cmb(\veclb') L(\veclb,\veclb')  \right)
	\left( \intl{3'} \cmb(\veclc') L(\veclc,\veclc')  \right) \Big>  \, .
\label{eqn:bi}
\end{eqnarray}
All these terms, and the necessary permutations, involve
an expectation value of three primary temperature anisotropies.
Under our assumption of Gaussian primary temperature fluctuations, such
expectation values vanish and thus there is no
contribution to the bispectrum or the skewness.

The trispectrum due to lensing alone can be calculated in a similar fashion. 
Introducing the power spectrum of lensing potentials, following
Refs.~\cite{Zal00,CooKes02}, we obtain the CMB trispectrum due to gravitational
lensing as:
\begin{eqnarray}    
\tilde T^{\cmb}(\bfl_1,\bfl_2,\bfl_3,\bfl_4) = -C_{l_3}^{\cmb} C_{l_4}^{\cmb}
\left\{ 
C^\pp_{|\vecl_1+\vecl_3|} \left[ (\vecla +\veclc) \cdot \veclc \right]
\left[ (\vecla + \veclc) \cdot \vecld \right] 
 + C^\pp_{|\vecl_2+\vecl_3|} \left[ (\veclb +\veclc) \cdot \veclc
\right] \left[ (\veclb +\veclc)  \cdot \vecld \right] \right\} 
+ \, {\rm Perm.} \, , 
\label{eqn:trilens}
\end{eqnarray}
where the permutations now contain 5 additional terms with the replacement of
$(l_3,l_4)$ by any other pair.

We can generalize our discussion of the power spectrum, bispectrum, and
trispectrum to that of the $n$-point correlation function in Fourier space.
In the absence of secondary anisotropies that correlate directly with the
lensing potential, the $n$-point correlation function will vanish for odd $n$
for the same reason that lensing alone did not generate a bispectrum.
All such terms would involve the expectation value of an odd number of
temperature fluctuations, and under the assumption of Gaussian primary
anisotropies, such expectation values must vanish.
This statement applies in particular to the case when measurements of
non-Gaussianity are made using CMB maps 
which have been cleaned {\it a priori} of secondary fluctuations using
information
such as the nonthermal frequency dependence of these fluctuations. We will
discuss the case of secondary anisotropies in the next subsection.

The lowest even $n$th correlator after the trispectrum is the six-point
correlation function in Fourier space. We can write the portion of the
connected part of this correlation function containing the lowest-order
contribution in  $\len$ as
\begin{eqnarray}
\left< \tilde \cmb(\bfl_1) \ldots
       \tilde \cmb(\bfl_6)\right>_c &=& \Big< \left( \cmb(\vecla) - \intl{1'} \cmb(\vecla')
L(\vecla,\vecla')\right) \ldots \left(\cmb(\veclc) - \intl{3'} \cmb(\veclc')
L(\veclc,\veclc')\right) \cmb(\vecld) \cmb(\vecle)  \cmb(\veclf) \Big>
+ \, {\rm Perm.} \nonumber \\
&& \quad =  -\Big<
\intl{1'} \cmb(\vecla') L(\vecla,\vecla') \ldots
\intl{3'} \cmb(\veclc') L(\veclc,\veclc') 
\cmb(\vecld) \cmb(\vecle) \cmb(\veclf) \Big> + \, {\rm Perm.} \nonumber \\
\label{eqn:tri}
\end{eqnarray}
Simplifying further, we see that the lowest order contribution in $\len$ thus
involves
\begin{eqnarray}
 \left< \tilde \cmb(\bfl_1) \ldots
       \tilde \cmb(\bfl_4)\right>_c 
&=&  C_{l_4}^\cmb C_{l_5}^\cmb  C_{l_6}^\cmb \Big<
\len(\vecla+\vecld) \len(\veclb+\vecle) \len(\veclc+\veclf) \Big>
\, \left[ (\vecla + \vecld) \cdot \vecld \right] \left[ (\veclb + \vecle) \cdot \vecle \right] \left[ (\veclc + \veclf) \cdot \veclf \right] + \, {\rm Perm.}
\end{eqnarray}
The connected part of the six-point correlation function in Fourier space is
thus proportional to the
bispectrum of lensing potentials. We can write
\begin{eqnarray}    
&&\tilde T_{6}^\cmb(\bfl_1,\bfl_2,\bfl_3,\bfl_4,\bfl_5,\bfl_6) = C_{l_4}^{\cmb} C_{l_5}^{\cmb} C_{l_6}^{\cmb} \Big[ 
B^\len(\vecl_1+\vecl_4,\vecl_2+\vecl_5,\vecl_3+\vecl_6) \left[ (\vecla +\vecld) \cdot \vecld \right] \left[ (\veclb + \vecle) \cdot \vecle \right]
\left[ (\veclc + \veclf) \cdot \veclf \right] \Big]\, + {\rm Perm.} \, 
\label{eqn:sixlens}
\end{eqnarray}
There are in total 120 such terms appearing in the six-point correlator when
we include all permutations, coming from the 20 different triplets
$(l_i,l_j,l_k)$ and the 6 permutations of each triplet.

We can generalize these derivations to the $n$-point temperature correlation in Fourier space under gravitational
lensing. In the following, note that contributions to $n$-point temperature
correlations in Fourier space come
from $(n/2)$-point correlations in the lensing potential. We can thus write the
connected part of the $n$-point temperature correlator, when $n > 2$, as
\begin{eqnarray}    
&&\tilde T_n^\cmb(\bfl_1,\ldots,\bfl_n) = \nonumber \\
&& \quad C_{l_{n/2+1}}^{\cmb} \ldots C_{l_n}^{\cmb} \Big[ 
T_{n/2}^\len(\vecl_1+\vecl_{(n/2)+1},\ldots,\vecl_{n/2}+\vecl_n) (\vecl_1 +\vecl_{(n/2)+1}) \cdot \vecl_{(n/2)+1} 
\ldots(\vecl_{n/2} + \vecl_n) \cdot \vecl_n \Big]\, + {\rm Perm.} \, , 
\label{eqn:nlens}
\end{eqnarray}
where $T_n^\len(\vecl_1,\ldots,\vecl_n)$ is the $n$-point correlator of the
lensing potential in Fourier space.
The permutations here now involve $n!/(n/2)!$ terms corresponding
to the replacement of $(l_{(n/2)+1},...,l_n)$ with one of the other
$n!/[(n/2)!(n/2)!]$
combinations and the $(n/2)!$ permutations of each combination.
As we have discussed, note that $\tilde T_n^\cmb(\bfl_1,\ldots,\bfl_n)=0$ when
$n$ is odd.

In the limit that the lensing potentials are Gaussian distributed,
$T_n^\len(\vecl_1,\ldots,\vecl_n)=0$ when $n>2$. Thus, lensing of CMB
anisotropies can only generate a trispectrum and, with smoothing, a kurtosis. 
The non-Gaussianity associated with the large-scale structure, however, will
induce non-Gaussian
contributions to the distribution of projected potentials such that 
$T_n^\len(\vecl_1,\ldots,\vecl_n) \neq 0$
for some $n$. Since large-scale structure most efficiently lenses the CMB at
redshifts close to 3,  where the non-Gaussianity is mild, we ignore the higher-order
correlations of lensing potentials and only consider the dominant power
spectrum, $C_l^\pp$, which contributes to
the trispectrum only.

Although theoretical predictions are made in terms of
ensemble-averaged correlation functions,
observationally we have access to only one realization of the CMB and one
realization of the large-scale structure.  The arbitrariness of the observed
realization of the large-scale structure induces additional cosmic variance
beyond that normally associated with the surface
of last-scatter.  One consequence is that when measured on a small patch of
the sky, the observed two-point correlation function of the lensed map is more
anisotropic than that of the unlensed map,
though isotropy holds when a sufficiently large region of the sky is
considered.  The excess anisotropy is induced by cosmic shear, and allows us to
reconstruct the lensing deflection angle from quadratic maps involving the CMB
temperature and polarization \cite{CooKes02}.  While we emphasize
one-point statistics in this paper, a more detailed account of how higher-order
statistics probe the local anisotropy induced by lensing may prove fruitful in
the future.

\subsection{Lensing-Secondary Correlations}
\label{SS:sec}

The above discussion applies to the case where other secondary fluctuations do
not contribute to 
temperature anisotropies. In practice, such a situation can be achieved when thermal CMB fluctuations are
separated from dominant secondary effects like the SZ contribution. In experiments where this is not possible,
say due to a lack of multifrequency data, additional non-Gaussianities will be present in
the CMB map due to correlations
between lensing potentials and the secondary anisotropies. The most significant of these contributions is to the three-point correlation function.
We can calculate this by replacing the $\tilde \cmb(\vecl)$ terms in
Eq.~(\ref{eqn:bi}) with $\cmb^t(\vecl)$.  By assumption, Gaussian instrumental noise cannot
generate a bispectrum, and as shown above neither does lensing alone.  The
total observed bispectrum is therefore that due to lensing-secondary
correlations \cite{SpeGol99,Zal00}
\begin{eqnarray} \label{E:SZbi}
&& B^{\cmb \tot}(\bfl_1,\bfl_2,\bfl_3) = -C_{l_1}^{\len\s} \left[C_{l_2}^\cmb (\veclb \cdot \vecla)  +C_{l_3}^\cmb 
(\veclc \cdot \vecla)\right] + \,  {\rm Perm.} \, ,
\end{eqnarray}
where permutations involve two additional terms with the replacement of $l_1$ with $l_2$ and $l_3$.
Here, $C_{l_1}^{\len\s}$ is the power spectrum describing correlations between
secondary anisotropies and the lensing potential generated by large-scale structure. These correlations were
discussed in detail in Ref.~\cite{SpeGol99} where it was found that the most significant correlation is the
one between lensing potentials and the SZ effect. We will use this correlation in illustrating our results.

The presence of secondary effects also modifies the 
trispectrum and generates an
additional contribution beyond the one discussed in Eq.~(\ref{eqn:trilens}). Following Ref. \cite{CooKes02},
we can write this contribution as
\begin{eqnarray}
&& T^{\cmb \s}(\bfl_1,\bfl_2,\bfl_3,\bfl_4) = C_{l_3}^{\len\s} C_{l_4}^{\len\s} \Big\{  C^\cmb_{l_1} (\veclc \cdot \vecla) (\vecld \cdot \vecla) +
 C^\cmb_{l_2} (\veclc \cdot \veclb) (\vecld \cdot \veclb) \nonumber \\
&&\quad + \left[ \veclc \cdot (\vecla + \veclc) \right]
\left[ \vecld \cdot (\veclb +\vecld) \right] C^\cmb_{|\vecl_1 + \vecl_3|}
+\left[ \vecld \cdot (\vecla + \vecld) \right]
\left[ \veclc \cdot (\veclb +\veclc) \right] C^\cmb_{|\vecl_1 + \vecl_4|}
\Big\} \nonumber \\
&& \quad \quad + \, {\rm Perm.} \,
\label{eqn:trisecfinal}
\end{eqnarray}
where permutations involve five additional terms involving the pairings of $(l_3,l_4)$.

Due to an increase in terms as one goes to higher order, we failed to obtain a general 
expression for the $n$-point correlator of temperature fluctuations in Fourier space due to
lensing-secondary correlations. As we will soon discuss, cumulants 
beyond the skewness are unlikely to be important as we find kurtosis to be undetectable even for
a perfect experiment with no noise and all-sky observations. We expect this to hold true even
when considering higher-order moments beyond the kurtosis.

\section{Skewness and Kurtosis}
\label{sec:S&K}

A simple way to identify the non-Gaussianity induced in the CMB by
gravitational
lensing is to measure the higher-order cumulants of its one-point probability
distribution function $P_{\rm{obs}}(\cmb^\tot; \theta)$ smoothed with beamwidth
$\theta$.  This observed one-point probability distribution function (PDF) is
actually a convolution of the signal PDF $P_{\rm{sig}}(\cmb^{\rm{sig}};
\theta)$
with the noise PDF $P_{\rm{noise}}(\cmb^\n; \theta)$ as described below, where
$\cmb^{\rm{sig}} = \tilde \cmb + \cmb^\s$ is the total of both lensed primary
and secondary contributions to the signal.  The signal PDF can be expressed in
terms of its cumulants, which we now proceed to calculate.
The third and fourth cumulants are proportional to
the dimensionless quantities known as the skewness, $S$, and the kurtosis, $K$,
respectively:
\begin{eqnarray}
S(\theta) &\equiv& \left[ \sigma(\theta) \right]^{-3} \int (\cmb^{\rm{sig}})^3
P_{\rm{sig}}(\cmb^{\rm{sig}}; \theta) \, d\cmb^{\rm{sig}}  \, , \nonumber \\
K(\theta) &\equiv& \left[ \sigma(\theta) \right]^{-4} \int (\cmb^{\rm{sig}})^4
P_{\rm{sig}}(\cmb^{\rm{sig}}; \theta) \, d\cmb^{\rm{sig}} - 3 \, .
 \nonumber \\
\end{eqnarray}
They can be expressed as integrals over the bispectrum and trispectrum derived
in the preceding Section according to Eq.~(\ref{eqn:mom}):
\begin{eqnarray} \label{E:SBKT}
S(\theta)  &=& \int \frac{d^2\bfl_1}{(2\pi)^2} \frac{d^2\bfl_2}{(2\pi)^2}
\frac{d^2\bfl_3}{(2\pi)^2}
(2\pi)^2 \delta_\dirac(\bfl_{123}) B^\tot(\bfl_1,\bfl_2,\bfl_3)
W(l_1 \theta) W(l_2 \theta) W(l_3 \theta) \, , \nonumber \\
K(\theta) 
&=& \int \frac{d^2\bfl_1}{(2\pi)^2} \frac{d^2\bfl_2}{(2\pi)^2}
\frac{d^2\bfl_3}{(2\pi)^2} \frac{d^2\bfl_4}{(2\pi)^2}
(2\pi)^2 \delta_\dirac(\bfl_{1234}) T^\tot(\bfl_1,\bfl_2,\bfl_3, \bfl_4)
W(l_1 \theta) W(l_2 \theta) W(l_3 \theta) W(l_4 \theta) \, . \nonumber \\
\end{eqnarray}
Inserting Eqs.~(\ref{E:SZbi}), (\ref{eqn:trilens}), and
(\ref{eqn:trisecfinal}) into the above
expressions, and adopting a Gaussian window function
$W(l \theta) = e^{-(l \sigma_b)^2/2}$ with $\sigma_b = \theta/\sqrt{8 \ln 2}$,
we obtain:
\begin{eqnarray} \label{E:Ktheta}
S(\theta)  &=& \frac{6}{(2\pi)^2 [\sigma(\theta)]^3} \int l_{1}^2 \, dl_1 \,
l_{2}^2 \, dl_2 \, C_{l_1}^{\cmb} C_{l_2}^{\len\s} I_1(\sigma_{b}^2 l_1 l_2)
\, e^{-\sigma_{b}^2 (l_{1}^2 + l_{2}^2)} \, ,
\nonumber \\
K^\pp(\theta) &=& \frac{12}{(2\pi)^3 [\sigma(\theta)]^4} \int \, dl_1
\, l_{1}^3
\, C_{l_1}^{\pp} e^{-\sigma_{b}^2 l_{1}^2} \left[ \int \, dl_2 \, l_{2}^2 \,
C_{l_2}^{\cmb} I_1(\sigma_{b}^2 l_1 l_2) \, e^{-\sigma_{b}^2 l_{2}^2} \right]^2
\, , \nonumber \\
K^{\len\s}(\theta) &=& \frac{12}{(2\pi)^3 [\sigma(\theta)]^4} \int \, dl_1 \,
l_{1}^3 \, C_{l_1}^{\cmb} e^{-\sigma_{b}^2 l_{1}^2} \left\{
\left[ \int \, dl_2 \, l_{2}^2 \,
C_{l_2}^{\len\s} I_1(\sigma_{b}^2 l_1 l_2) \, e^{-\sigma_{b}^2 l_{2}^2}
\right]^2 \right. \nonumber \\
&& \left. - \frac{1}{2\pi} \int l_{2}^2 \, dl_2 \, l_{3}^2 \, dl_3 \, d\varphi
\, C_{l_2}^{\len\s} C_{l_3}^{\len\s} I_1\left(\sigma_{b}^2 l_2
\sqrt{l_{1}^2 + l_{3}^2 +2 l_1 l_3 \cos \varphi}\right)
\, e^{-\sigma_{b}^2 (l_{2}^2 + l_{3}^2 + l_1 l_3 \cos \varphi)}
\frac{l_1 \cos \varphi + l_3 \cos^2 \varphi}{\sqrt{l_{1}^2 + l_{3}^2 +2 l_1 l_3
\cos \varphi}}
\right\} \, . \nonumber \\
\end{eqnarray}
Here $I_1(x)$ is a modified Bessel function of the first kind.

In the presence of instrumental noise, the observed one-point probability
distribution function (PDF) will be a convolution of the signal PDF
characterized by the skewness and kurtosis given above, and a Gaussian noise
PDF: $P_{\rm{obs}}(\cmb^\tot) = \int d\tau \, P_{\rm{sig}}(\tau)
P_{\rm{noise}}(\cmb^\tot - \tau)$.
In order to perform this convolution we must first determine the explicit
form of the signal PDF that will have nonzero skewness or kurtosis, but
vanishing higher cumulants.  To do this, we follow the formalism discussed in
Ref.~\cite{Edge} and references therein.  The PDF of a random variable $\delta$
with zero mean and variance $\sigma^2$ can be expressed as a Gram-Charlier
series in the normalized variable $\nu \equiv \delta/\sigma$:
\begin{equation} \label{E:GCS}
p(\nu) = c_0 \phi(\nu) + \frac{c_1}{1!} \phi^{(1)}(\nu) +
\frac{c_2}{2!} \phi^{(2)}(\nu) + \ldots,
\end{equation}
where $\phi(\nu) \equiv (2\pi)^{-1/2} e^{-\nu^2/2}$ is a Gaussian
distribution.  The
$\phi^{(l)}(\nu)$ are derivatives of the Gaussian distribution
with respect to $\nu$:
\begin{equation}
\phi^{(l)}(\nu) \equiv \frac{d^l \phi}{d\nu^l} = (-1)^l H_l(\nu) \phi(\nu),
\end{equation}
and the $H_l(\nu)$ are Hermite polynomials with the unconventional
normalization,
\begin{equation} \label{E:orthog}
\int_{-\infty}^{\infty} H_l(\nu) H_m(\nu) \phi(\nu) d\nu = l! \, \delta_{lm}
\, .
\end{equation}
The central moments of the PDF are defined as
\begin{equation}
\mu_l \equiv \sigma^l \int_{-\infty}^{\infty} p(\nu) \nu^l \, d\nu \, ,
\end{equation}
while the cumulants or ``connected'' portions of these moments can be derived
from the relation
\begin{equation} \label{E:cumu}
M_l \equiv \frac{d^l \ln \langle  e^{t \delta} \rangle }{dt^l} \, .
\end{equation}
Using the expansion (\ref{E:GCS}) and the orthogonality relation
(\ref{E:orthog}), the coefficients of the Gram-Charlier series can be
expressed in terms of the central moments.  By inverting Eq.~(\ref{E:cumu}), the
central moments can then be reexpressed in terms of cumulants.  As discussed
in the previous Section, the assumption that the lensing potential is
Gaussian implies that all cumulants of higher order than the kurtosis must
vanish.  Using this result, we can rewrite the Gram-Charlier expansion as a
power series in the skewness $S$ or kurtosis $K$, which in the case of lensing
will be small quantities.
\begin{eqnarray} \label{E:PDFs}
p(\nu) &=& \left[1 + \frac{1}{3!}S H_3(\nu) + \frac{10}{6!}S^2 H_6(\nu)
+ \ldots \right] \phi(\nu) \nonumber \, , \\
p(\nu) &=& \left[1 + \frac{1}{4!}K H_4(\nu) + \frac{35}{8!}K^2 H_8(\nu)
+ \ldots \right] \phi(\nu) \, . \nonumber \\
\end{eqnarray}
These power series can be convolved with a Gaussian noise PDF of variance
$\sigma_{\rm{noise}}^2(\theta)$ to obtain the observed PDF
$P_{\rm{obs}}(\cmb^\tot)$.
To linear order in the true skewness and kurtosis, we find:
\begin{eqnarray}
S_{\rm{obs}}(\theta) &=& S(\theta) \left\{
\frac{\sigma^2(\theta)}{\sigma^2(\theta) +
\sigma_{\rm{noise}}^2(\theta)} \right\}^{3/2} \, , \nonumber \\
K_{\rm{obs}}(\theta) &=& \frac{K(\theta)}{8} \left\{
\frac{5\sigma^4(\theta) - 3\sigma_{\rm{noise}}^4(\theta)
- 6\sigma^2(\theta) \sigma_{\rm{noise}}^2(\theta)}{\left[ \sigma^2(\theta) +
\sigma_{\rm{noise}}^2(\theta) \right]^2} + 3 \right\} \, . \nonumber \\
\end{eqnarray}
As expected, the observed skewness and kurtosis converge to the signal values
in the absence of noise and to zero in the case when  the Gaussian noise is
dominant.  To actually observe skewness or kurtosis in an experimental sky map,
we must construct estimators for these quantities using our data points, the
$N = 4\pi f_{\rm{sky}}/\pi(\theta/2)^2$ pixels in the map.
We can write estimators for the skewness and kurtosis as
\begin{eqnarray} \label{E:est}
\widehat{S \sigma^3} &\equiv& \frac{1}{N} \sum_{i=1}^{N} (x_i - \bar{x})^3 \, ,
\nonumber \\
\widehat{K \sigma^4} &\equiv& \frac{1}{N} \sum_{i=1}^{N} (x_i - \bar{x})^4 - 3
\left[ \frac{1}{N} \sum_{i=1}^{N} (x_i - \bar{x})^2 \right]^2
\, , \nonumber \\
\end{eqnarray}
where $\bar{x} = \frac{1}{N} \sum_{i=1}^{N} x_i$ is the traditional estimator
for the mean of a distribution.  For a distribution like that of the CMB
anisotropies which is {\it a priori} defined to have a zero mean, we find:
\begin{eqnarray}
\langle \widehat{S \sigma^3} \rangle &=& (1 - \frac{3}{N} + \frac{2}{N^2})
S \sigma^3 \, , \nonumber \\
\langle \widehat{K \sigma^4} \rangle &=& (1 - \frac{7}{N} + \frac{12}{N^2}
- \frac{6}{N^3}) K \sigma^4 - \frac{6}{N}(1 - \frac{1}{N}) \sigma^4\, .
\nonumber \\
\end{eqnarray}
These are biased estimators, as has been noted elsewhere under a different
context \cite{HuiGaz99},
but in the large-$N$ limit they converge to the
desired quantities.  Assuming that the underlying PDF is Gaussian, the variance
of these estimators to lowest order in $1/N$ is given by:
\begin{eqnarray} \label{E:exvar}
\sigma_{\widehat{S \sigma^3}}^2 &\equiv& \langle (\widehat{S \sigma^3})^2
\rangle - \langle \widehat{S \sigma^3} \rangle^2 = \frac{3!}{N} \sigma^3 \, ,
\nonumber \\
\sigma_{\widehat{K \sigma^4}}^2 &\equiv& \langle (\widehat{K \sigma^4})^2
\rangle - \langle \widehat{K \sigma^4} \rangle^2 = \frac{4!}{N} \sigma^4 \, .
\nonumber \\
\end{eqnarray}

An alternate derivation of these variances can be obtained from the explicit
form of the PDFs following Eq.~(\ref{E:PDFs}).  If $N$ pixels or
data points are collected and binned
such that $p_i$ is the probability that a data point will fall within bin $i$
and $\sigma_i$ is the standard deviation of that probability,
then the best variance of a parameter $\epsilon$ characterizing the PDF is
given by the Cram\'er-Rao bound \cite{KenStu69},
\begin{equation} 
\label{E:disc}
\frac{1}{\sigma_{\epsilon}^2} = \sum_i \left( \frac{\partial p_i}{\partial
\epsilon} \right)^2 \frac{1}{\sigma_{i}^2} \, .
\end{equation}
If the error on each bin is assumed to be Poisson, then $\sigma_{i}^2 = p_i/N$.
In the limit of a continuous PDF, $p_i \to p(\nu) \, d\nu$ and
the discrete sum (\ref{E:disc}) becomes an integral:
\begin{equation} \label{E:cont}
\frac{1}{\sigma_{\epsilon}^2} = N \int \left( \frac{\partial p}{\partial
\epsilon} \right)^2 p^{-1} \, d\nu \, .
\end{equation}
Inserting Eq.~(\ref{E:PDFs}) into Eq.~(\ref{E:cont}) under the Gaussian null hypothesis
$S = K = 0$, we find lowest attainable errors as $\sigma_{\epsilon}^2 = 3!/N, 4!/N$ for 
$\epsilon = S, K$ in agreement with the explicit calculation of the variance
of our estimators noted in Eq.~(\ref{E:exvar}).  Further discussion of the
variance associated with different estimators for the skewness and kurtosis
is included in the Appendix.

\begin{figure}[t]
\centerline{
\psfig{file=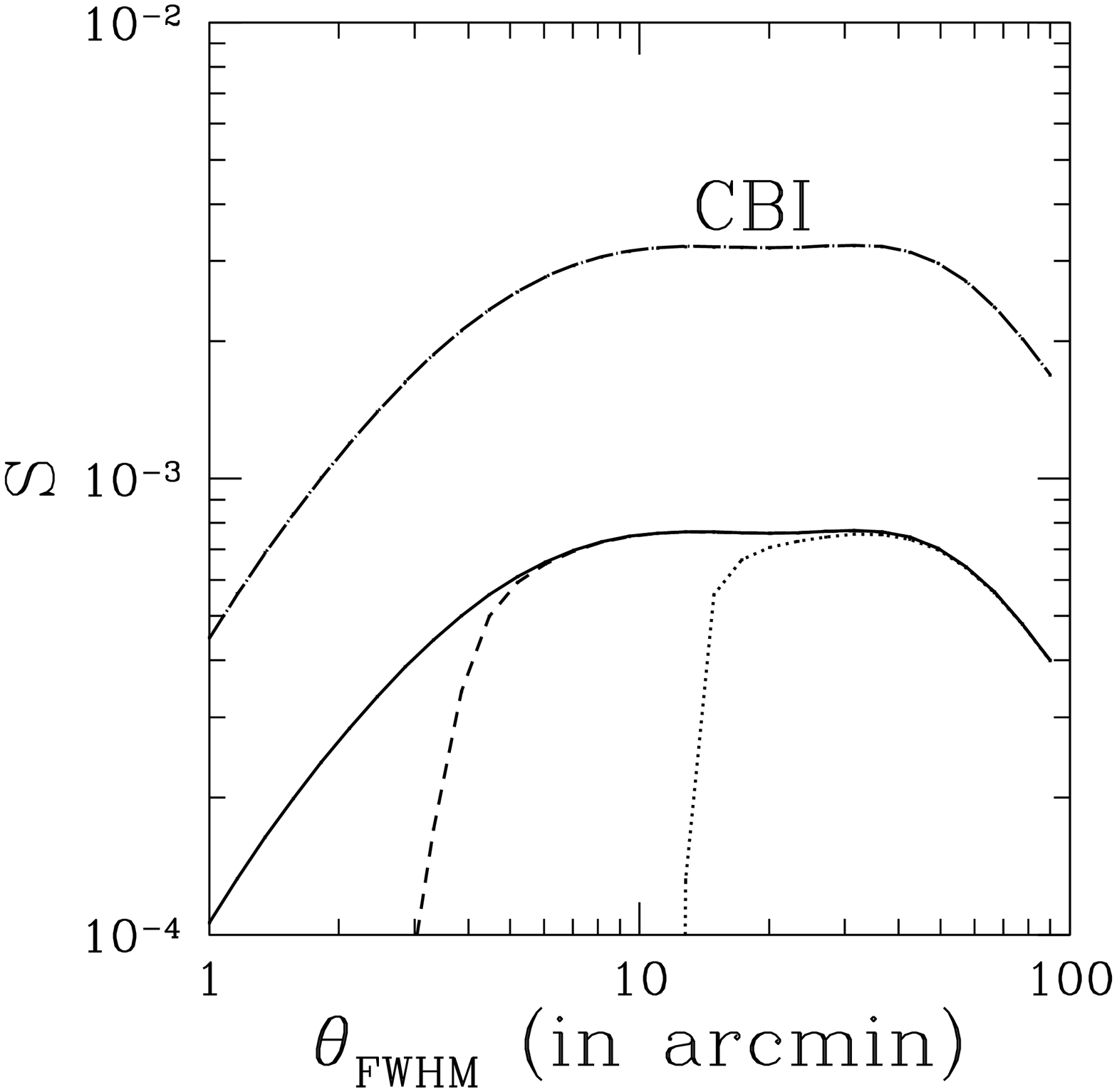,width=3.5in,angle=0}
\psfig{file=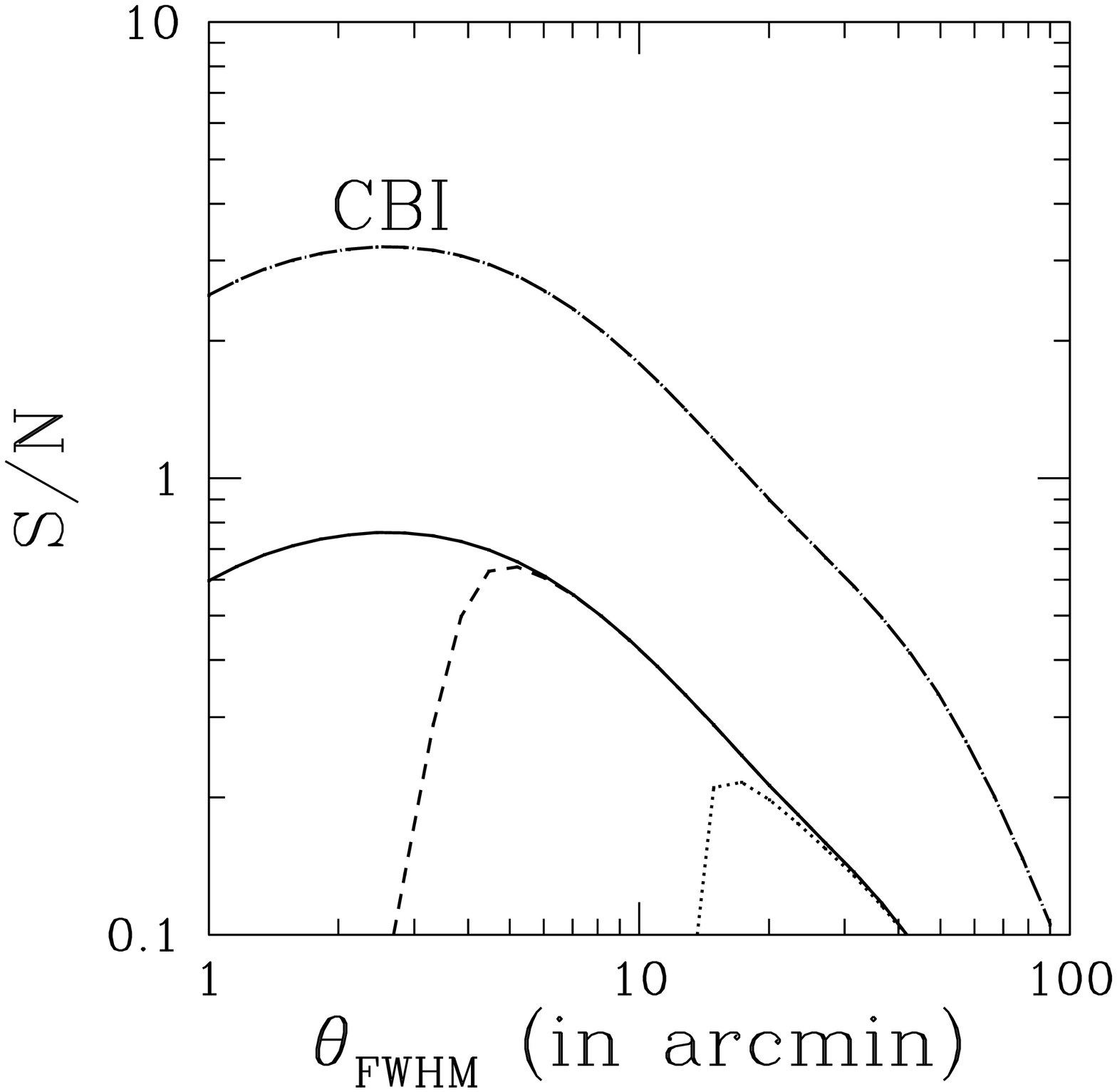,width=3.5in,angle=0}
}
\caption{{\it Left:} The skewness due to lensing-SZ correlations for a perfect
(no-noise) experiment (solid line),
Planck (dashed line), and MAP (dotted line) for $\sigma_8 = 0.9$. The CBI
$1 \sigma$ upper bound of $\sigma_8 \leq 1.2$ leads to a higher value for
the skewness as indicated by the dot-dashed line.
{\it Right:} The signal-to-noise ratio for the detection of skewness in
CMB data with curves labeled as in the left figure. We assume full
sky-coverage; for partial sky coverage the
signal-to-noise ratio scales as $\sqrt{f_{\rm sky}}$, where $f_{\rm sky}$ is the fraction of sky covered.}
\label{fig:skew}
\end{figure}

\section{Results \& Discussion}
\label{sec:results}

\subsection{Skewness}

We illustrate in Fig.~(\ref{fig:skew}) our results for skewness due to the
correlation between lensing and 
the SZ effect.  We calculate this correlation following Ref.~\cite{Coo01} using
the halo approach to  
large-scale structure \cite{CooShe02}. 
The skewness approaches zero at small values of the smoothing scale, consistent
with our conclusion  that no non-Gaussian signatures exist in the PDF in the
limit of infinite resolution.
As shown, skewness due to the lensing-SZ
correlation peaks at an angular scale of tens of arcminutes, which is in the
range of interest to upcoming experiments
such as MAP and Planck.  When calculating expected signal-to-noise ratios for
these experiments, we use detector sensitivities and resolutions tabulated in
Ref. \cite{Cooetal00}. For simplicity, we 
combine information from individual frequency channels to form one estimate
of temperature with an overall noise given by
inversely weighting individual noise contributions.

The skewness as shown has signal-to-noise ratios slightly less than
unity suggesting that its detection may be hard and
potentially affected by noise.  However, recent small-scale excess-power
detections by experiments such as CBI \cite{Masetal02} raise the possibility
that we may have underestimated the lensing-SZ correlation and thus the
skewness.  The lensing-SZ power spectrum $C_l^{\len\s}$ is roughly proportional to the
fifth power of $\sigma_8$, the standard deviation of linear mass fluctuations
within an $8 h^{-1}$ Mpc sphere.  If we adopt the CBI $1\sigma$ upper bound
of $\sigma_8 \leq  1.2$ \cite{Masetal02} as opposed to the value $\sigma_8 = 0.9$ suggested
by previous studies, our signal increases by a factor of 4.21.
In this case, Planck could conceivably detect skewness with a signal-to-noise
of 2.5.
The potential for detection of the temperature skewness is consistent with 
previous expectations that the temperature anisotropy bispectrum due to lensing-SZ correlation can be detected in future data \cite{SpeGol99}.
The cumulative signal-to-noise for skewness, however, is
significantly smaller than that for the full bispectrum because the skewness
is a single number while the bispectrum contains
all information related to non-Gaussianities at the three-point level.
As described below,
we find a similar reduction in signal-to noise for kurtosis when compared to
the full trispectrum.

The frequency dependence of the SZ effect allows us to construct an SZ map of
the sky as well as a temperature map with the SZ effect removed.  This provides
us a unique opportunity to test our understanding of non-Gaussianity at the
three-point level.  If skewness is purely a consequence of lensing-SZ
correlations as posited in this paper, then the skewness obtained by
combining one measurement of the SZ map with two measurements of the SZ-cleaned
temperature map at the same location using the estimator in Eq.~(\ref{E:est})
should be precisely one third that produced by three measurements of the total
anisotropy map.  This corresponds to the fact that the composite map will
sample only one of the three permutations appearing in Eq.~(\ref{E:SZbi}).

\begin{figure}[t]
\centerline{
\psfig{file=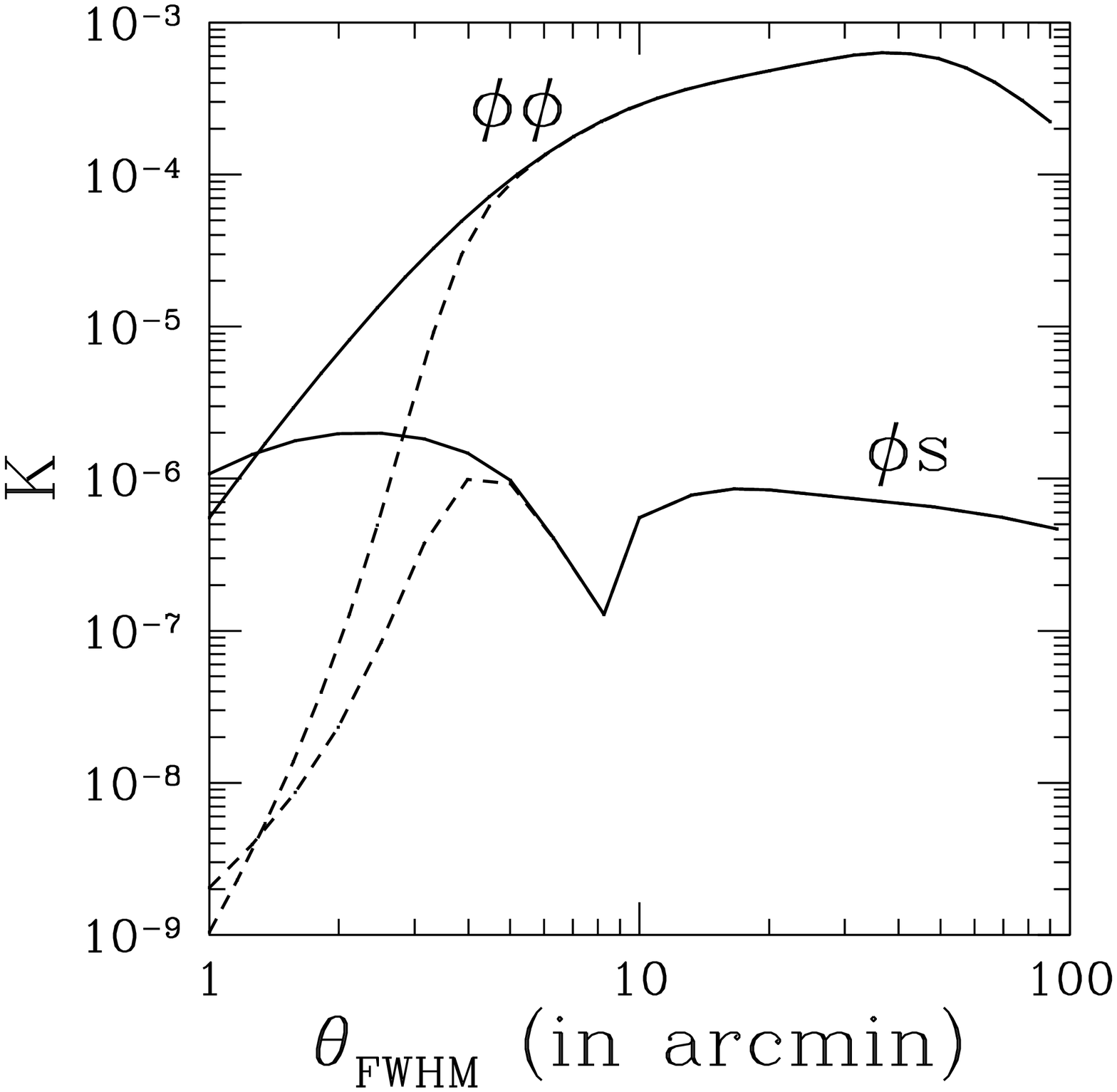,width=3.5in,angle=0}
\psfig{file=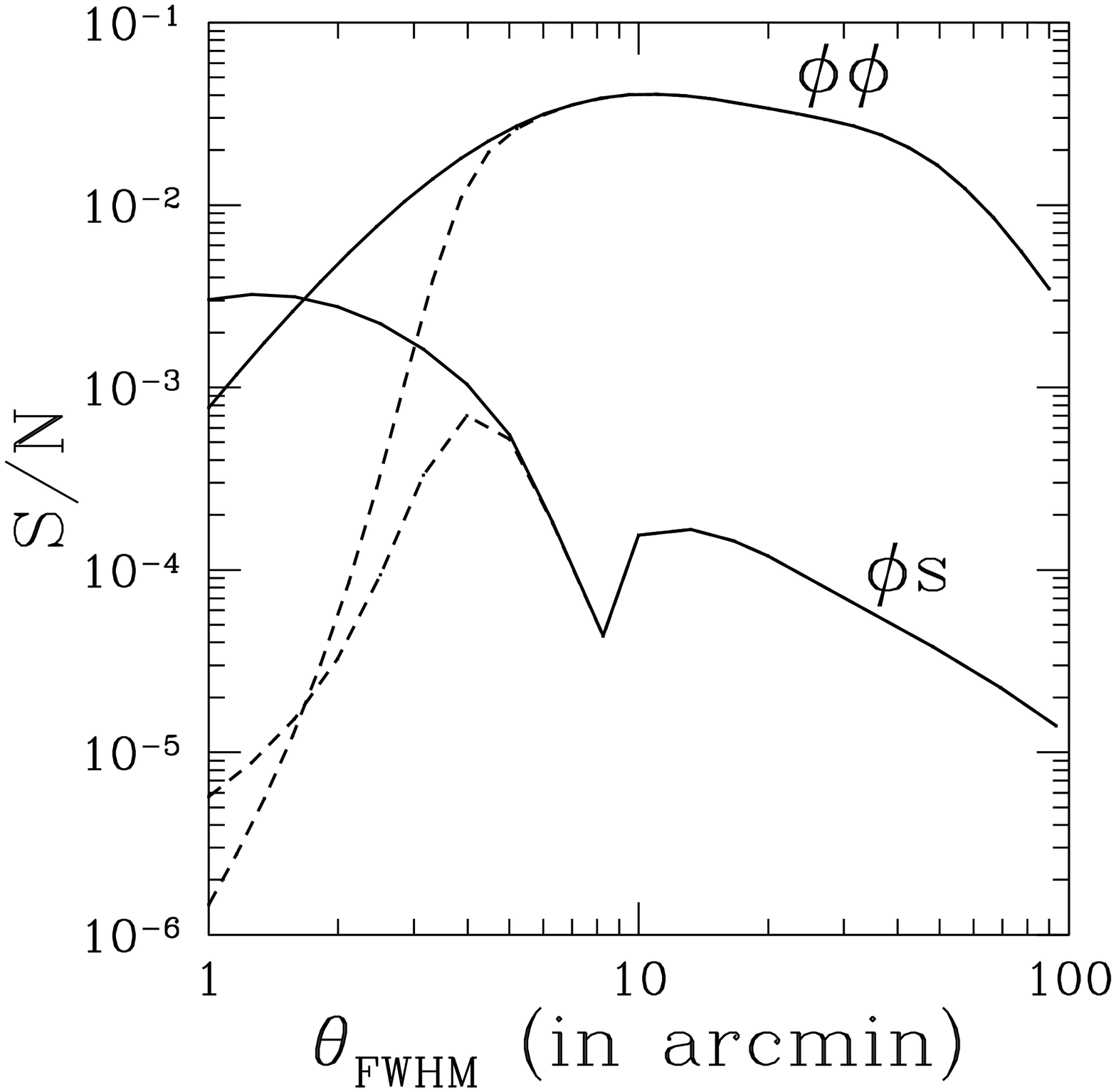,width=3.5in,angle=0}
}
\caption{{\it Left:} The kurtosis $K^\pp$ due to lensing autocorrelations and
$K^{\len\s}$ due to lensing-SZ cross-correlations for a perfect (no-noise)
experiment (solid line) and Planck (dashed line). The kurtosis due to
lensing-SZ correlations is negative at smoothing scales below the kink at
$\sim$ 8 arcminutes and positive thereafter; its absolute value is shown here.
{\it Right:} The signal-to-noise ratio for the detection of kurtosis in CMB
data with curves labeled as in the left figure.  We assume full sky-coverage;
for partial sky coverage the signal-to-noise ratio scales as
$\sqrt{f_{\rm sky}}$, where $f_{\rm sky}$ is the fraction of sky covered.}
\label{fig:kurt}
\end{figure}

\subsection{Kurtosis}

Both lensing kurtosis $K^\pp$ and the kurtosis $K^{\len\s}$ due to lensing-SZ
correlations are undetectable even for a perfect no-noise experiment as
illustrated in Fig.~(\ref{fig:kurt}).  Since the cumulative signal-to-noise
ratio for $K^{\len\s}$ is well below one, we expect it to remain undetectable
despite any uncertainty in our calculation of the SZ effect. Note our
prediction of the lensing kurtosis $K^\pp$ is likely to be more certain since
it only depends on the matter power spectrum, with contributions
coming mainly from the linear regime. Thus, uncertainties in non-linear aspects
of clustering are unlikely to affect our conclusion.

The signal-to-noise value for $K^\pp$
can be compared to the cumulative signal-to-noise ratio for the direct
detection of the full trispectrum due to lensing, which in the case of Planck
can be as high as $\sim$ 55 \cite{Zal00}.
Consequently, although the lensing kurtosis cannot be detected directly
from the data, lensing effects associated with this kurtosis can be used to
reconstruct the lensing deflection angle as described in
Refs.~\cite{Hu01b,CooKes02}, again with cumulative signal-to-noise ratios
significantly greater than that for the kurtosis itself.
The higher signal-to-noise ratio in lensing reconstruction is possible for
two reasons.  Unlike the kurtosis, which averages indiscriminately over all
configurations of the trispectrum as shown in Eq.~(\ref{E:SBKT}), lensing
reconstruction is sensitive to certain configurations of the
trispectrum, mainly those that contribute to the power spectrum of squared
temperature.
This avoids severe positive-negative cancellations that significantly reduce
the signature of non-Gaussianity.
Secondly, the noise contribution associated with lensing reconstruction is also
{\it a priori} reduced through a filter which is designed to
extract information on the lensing potentials
optimally.

The low signal-to-noise associated with the kurtosis is also consistent with
the fact that real-space moments, in general, suffer from excess noise. Though
such statistics are easily measurable in data, they do not provide the most
optimal methods to search for the existence of non-Gaussian signatures.
While we recommend construction of cumulants such as skewness and kurtosis as a
first step in understanding
non-Gaussianity from effects such as lensing, we suggest that full
measures of quantities such as bispectrum and trispectrum will  be necessary to
fully understand the
non-Gaussian behavior of lensing. If measurement of such statistics are still
cumbersome, 
we suggest the use of quadratic statistics in real space, such as the squared-temperature--temperature 
\cite{Coo02a} and the squared-temperature--squared-temperature \cite{CooKes02}
power spectra which probe certain configurations of the bispectrum and
trispectrum.

\acknowledgments

This work was supported in part by NSF
AST-0096023, NASA NAG5-8506, and DoE DE-FG03-92-ER40701.  Kesden
acknowledges the support of an NSF Graduate Fellowship, and AC
acknowledges support from the Sherman Fairchild Foundation.

\appendix
\section{Variance of Skewness and Kurtosis Estimators}

A question arose during the composition of this paper as to the appropriate
variance for estimates of the skewness and kurtosis of a Gaussian distribution.
The true skewness and kurtosis of a Gaussian distribution are necessarily
zero, but given $N$ data points $x_i$ drawn from this distribution even
unbiased estimators will yield results distributed about zero with some
variance.  Some sources (e.g., \cite{NumRec})
indicate variances of $15/N$ and $96/N$ respectively for
the skewness and kurtosis estimators defined in Eq.~(\ref{E:est}) as opposed to
our values of $6/N$ and $24/N$.
This discrepancy prompted us to investigate further.  The estimators
of Eq.~(\ref{E:est}) differ from those given in Ref.~\cite{NumRec} in that
they are estimators for the third and fourth cumulants rather than the
dimensionless skewness and kurtosis to which they are proportional.
Assuming an underlying Gaussian distribution with a variance of unity,
standard propagation of errors reveals that the two pairs of estimators have
the same variances to lowest order in $1/N$.
However, the na\"{\i}ve estimators
\begin{eqnarray} \label{E:naiveest}
\widehat{S \sigma^3}^\prime &\equiv& \frac{1}{N} \sum_{i=1}^{N} x_{i}^3 \,  {\rm and}
\nonumber \\
\widehat{K \sigma^4}^\prime &\equiv& \frac{1}{N} \sum_{i=1}^{N} x_{i}^4 - 3
\left[ \frac{1}{N} \sum_{i=1}^{N} x_{i}^2 \right]^2 \,  \nonumber \\
\end{eqnarray}
do indeed have variances of $15/N$ and $96/N$ for skewness and kurtosis
respectively.  We show this explicitly for the
na\"{\i}ve skewness estimator $\widehat{S \sigma^3}^\prime$.  The ensemble
average of this estimator is simply $S \sigma^3$ so it is truly an unbiased
estimator for the skewness.  However, taking the ensemble average of 
$(\widehat{S \sigma^3}^\prime)^2$ we find
\begin{eqnarray}
\langle (\widehat{S \sigma^3}^\prime)^2 \rangle = \frac{1}{N} \left[ \mu_6 +
(N - 1) S^2 \sigma^6 \right] \, ,
\end{eqnarray}
leading to a variance
\begin{eqnarray}
\sigma_{\widehat{S \sigma^3}^\prime}^2 \equiv \langle
(\widehat{S \sigma^3}^\prime)^2 \rangle - \langle \widehat{S \sigma^3}^\prime
\rangle^2 = \frac{1}{N} (\mu_6 - S^2 \sigma^6) \, .
\end{eqnarray}
For a Gaussian distribution, $\mu_6 = 15 \sigma^6$ and $S = 0$, implying that
this estimator measures skewness with a variance of $15/N$ and is therefore
less sensitive than $\widehat{S \sigma^3}$ defined in Eq.~(\ref{E:est}) which
was shown to have a variance of $6/N$.  An entirely analogous calculation
shows that the na\"{\i}ve kurtosis estimator in Eq.~(\ref{E:naiveest}) has a
variance of $96/N$, not $24/N$.

Why do the estimators of Eq.~(\ref{E:est}) outperform those of
Eq.~(\ref{E:naiveest})?  Although the true mean of the
underlying Gaussian distribution has been chosen to be zero, the estimated
mean $\bar{x} = \frac{1}{N} \sum_{i=1}^{N} x_i$ of $N$ data points will not
necessarily vanish.  The more sophisticated estimators of Eq.~(\ref{E:est})
take this into account by subtracting the estimated mean from each data point,
and are therefore able to provide lower-variance estimates of the skewness and
kurtosis.  These lower values for the variances are adopted for all results
concerning signal-to-noise mentioned in this paper.

\end{document}